\documentclass{l4dc2024}

\usepackage{comment,enumitem}
\usepackage{amsfonts}
\usepackage{mathtools,mathbbol}
\usepackage{multicol,capt-of}
\usepackage{hhline}
\usepackage{multirow}
\usepackage{algorithm}
\usepackage{graphicx}
\usepackage{textcomp}
\usepackage{xcolor,framed}
\usepackage{colortbl}
\usepackage{ragged2e,adjustbox}
\usepackage{bm,xr}
\allowdisplaybreaks
\newtheorem{assumption}{Assumption}

\makeatletter
\newcommand*{\addFileDependency}[1]{
\typeout{(#1)}
\@addtofilelist{#1}
\IfFileExists{#1}{}{\typeout{No file #1.}}
}\makeatother

\newcolumntype{C}[1]{>{\centering\arraybackslash}p{#1}}

\usepackage{tikz}
\usetikzlibrary{arrows,shapes,positioning,automata,backgrounds,petri}

\title[IOC as EIV]{Inverse Optimal Control as an Errors-in-Variables Problem}
\usepackage{times}

 \coltauthor{\Name{Rahel Rickenbach} \Email{rrahel@ethz.ch}\\
  \Name{Anna Scampicchio} \Email{ascampicc@ethz.ch}\\
  \Name{Melanie N. {Zeilinger}} \Email{mzeilinger@ethz.ch}\\
  \addr Institute for Dynamic Systems and Control, Leonhardstrasse 21, 8051 Zurich}

\begin{document}

\maketitle

\begin{abstract}%
Inverse optimal control (IOC) is about estimating an unknown objective of interest given its optimal control sequence. 
However, truly optimal demonstrations are often difficult to obtain,  
e.g., due to human errors or inaccurate measurements. This paper presents an IOC framework for objective estimation from multiple sub-optimal demonstrations in constrained environments. It builds upon the Karush-Kuhn-Tucker optimality conditions, and addresses the Errors-In-Variables problem that emerges from the use of sub-optimal data. The approach presented is applied to various systems in simulation, and consistency guarantees are provided for linear systems with zero mean additive noise, polytopic constraints, and objectives with quadratic features.
\end{abstract}


\section{Introduction}
Applications in robotics and control often involve complex and demanding tasks in constrained environments. While strategies such as Model Predictive Control \citep{kouvaritakis2016} have been successfully deployed to address these challenges, their performance depends heavily on the design of the objective function, which is often nontrivial. In fact, the translation of a complex goal description into a suitable objective is often unintuitive, and its tuning can be delicate. Thus, inverse optimal control (IOC) methods \citep{FengShunLin:2021,abazar2020}  provide a promising tool to tackle this issue: starting from a partially specified objective (e.g., described by a set of basis functions), they aim to estimate the missing parameters from optimal control sequences (also known as \textit{demonstrations}).  \\
However, available control sequences may be suboptimal in practice, as they are, e.g., provided by humans, or affected by noise.  The methods in \cite{englert2017} and \cite{menner2019}, which rely on the Karush-Kuhn-Tucker (KKT) conditions \citep{kuhn2014}, address these issues by allowing for a slight suboptimality in the demonstrations 
using a least-squares approximation of the stationarity condition. However, the estimate obtained is not consistent:
it does not deal with the errors entering the regressors due to the sub-optimality of the given controls -- in other words, the Errors-In-Variables (EIV) nature of the problem \citep{griliches1974} is not addressed. 
A possibility to address this issue consists in jointly estimating demonstrations and optimal objective parameters, as done by \cite{Hatz2012}.  The same idea is used in \cite{menner2020}, where a probabilistic model taking Gaussian measurement noises into account is considered.  However, the majority of IOC approaches rely on deterministic models and do not consider stochastic noise in the dynamics.  
Another exception is the approach in \cite{nakano2023inverse}, where a stochastic demonstration model is considered,  but knowledge of the optimal input sequence is still assumed.  
An intrinsic probabilistic view of the problem is given by the formulation in Inverse Reinforcement Learning (IRL), where data are modeled as Markov Decision Processes \citep{feinberg2012} and thus intrinsically deal with suboptimal demonstrations. Solutions are typically obtained with entropy maximization \citep{ziebart2008,finn2016}. 
However,  in contrast to IOC, the IRL formulation does not allow for a natural inclusion of constraints. 

\paragraph*{Contribution} This work presents an IOC framework capable of dealing with suboptimal and noisy demonstrations.  It starts from a formulation based on the KKT conditions and rephrases it in terms of an EIV problem \citep{griliches1974}: to the best of the authors' knowledge, this is novel in the IOC and IRL literature.  We present two approaches considering different assumptions to solve it: when the distribution of noises is known, we start from a Bayesian interpretation of the problem \citep{dellaportas1995} and rely on Markov Chain Monte Carlo (MCMC) \citep{gilks1995}, while when no such an information is given, we leverage total least squares \citep{golub1980}. Our approach relates to that in \cite{menner2020}, because it jointly estimates demonstration and optimal objective parameters working with probabilistic models; 
however, differently from that work, we provide principled ways of estimating hyper-parameters, and study the case in which disturbances are non-Gaussian. Our approaches not only allow for an improved estimation performance with respect to the state-of-the-art, as shown in the numerical tests, but also provide consistency guarantees in the case of linear systems with zero-mean additive noise, polytopic constraints and quadratic objectives.

\paragraph*{Notation}
We denote with $\mathbb{0}_{n,m}$ and $\mathbb{1}_{n,m}$ a matrix of dimension $n \times m$ filled with zeros and ones, respectively, and with $\mathbb{I}_n$ the $n-$dimensional identity matrix. 
Given a vector $a \in \mathbb{R}^{b}$ and a function \mbox{$h(a): \mathbb{R}^{b} \rightarrow \mathbb{R}^{c}$}, then $\nabla_{a}h(a)$ returns the Jacobian of $h(a)$ with respect to $a$. A Gaussian random vector $\mathfrak{a}$ with mean $\mu_{\mathfrak{a}}$ and covariance $\Sigma_{\mathfrak{a}}$ is given as $\mathcal{N}(\mathfrak{a}; \mu_{\mathfrak{a}}, \Sigma_{\mathfrak{a}})$, while an Inverse Wishart $\mathfrak{A}$ random matrix with scale matrix $W_{\mathfrak{A}}$ and $m_{\mathfrak{A}}$ degrees of freedom will be denoted as $IW(\mathfrak{A}; W_{\mathfrak{A}}, m_{\mathfrak{A}})$. 
\section{Problem Statement}
We introduce the IOC problem in Section \ref{subsec:optimalsolutionanditsforwardproblem}, and present in Section \ref{subsec:theinversekarushkuhntuckerappraoch} the solution strategy from \cite{englert2017} building upon the KKT conditions. This is followed by the definition of the sub-optimal demonstrations considered and of the least-squares approximation of the inverse KKT approach in Sections \ref{subsec:suboptimaldemonstrations} and~\ref{subsec:aleastsquareapproximationofinversekkt}, respectively.

\subsection{Forward and Inverse Optimal Control Problem}
\label{subsec:optimalsolutionanditsforwardproblem}
\noindent We consider a known, deterministic, discrete-time system
  $  x_{k+1} = f(x_k,u_k),$ 
where $x_k \in \mathbb{R}^n$ indicates the state and $u_k \in \mathbb{R}^m$ the input vector at time step $k$, generating state and input trajectories of length $N+1$ and $N$, respectively. For notational simplicity and a compact representation of the IOC problem, we stack the input variables over their respective horizon length in $U = ( u_0^{\top},\hdots,u_{N-1}^{\top})^{\top} \in \mathbb{R}^{Nm}$. As for the state $x_k$, we represent it through  the operator
\begin{equation}
    F_{k}(U,x_{0}) = 
    \begin{dcases}
        x_{0}  & \mathrm{if}  \ k = 0, \\[1ex]
        f(F_{k-1}(U,x_{0}),u_{k-1}) & \mathrm{if}  \ k \geq 1,
    \end{dcases}
    \label{eq:dynamics}
\end{equation}
and collect the sequence of all $F_{k}(U,x_{0})$ in \mbox{$\mathcal{F}_{U,x_0}=\{F_0(U,x_{0}), \hdots, F_{N}(U,x_{0}) \}$}. The optimal control problem considered is referred to as the \textit{forward problem} and, with horizon length~$N$ and $I$ known inequality constraints $\{g_i(\cdot,\cdot)\}_{i=1}^I$, reads as 
\begin{equation}
    \min_{U} \sum_{k = 0}^{N-1} \theta^{\top} \phi(F_{k}(U,x_{0}),u_k) \quad  \text{subject to } \begin{cases}
        &g_i(F_{k}(U,x_{0}),u_k) \leq 0 \\
    & x_0 = x(0) \\
    & k = 0,\hdots,N, \  i = 1,\hdots,I.
    \end{cases}
\label{eq:optimizationproblem}
\end{equation}
The objective is modeled as a linear combination of given features, collected in the vector $\phi(\cdot) \in \mathbb{R}^q$, with an unknown coefficient vector $\theta \in \mathbb{R}^q$.  We further assume that $f$, $g_i$ and $\phi$ are continuously differentiable for all \mbox{$i = 1,\hdots,I$}.   Ultimately, the \textit{inverse optimal control} problem follows as inferring the unknown parameter vector $\theta$ from the optimal input sequence, denoted with $U^*= (u_0^{*,\top}, \hdots, u_{N-1}^{*,\top})$. 

\subsection{The Inverse KKT Approach}
\label{subsec:theinversekarushkuhntuckerappraoch}
Considering optimal demonstrations, an estimate of the unknown parameter vector $\theta$ can be obtained based on the KKT conditions, which provide necessary conditions for the solution of an optimization problem. 
To present them, we introduce the Lagrangian multipliers \mbox{$\lambda_{i,k} \in \mathbb{R}$}, for all \mbox{$k = 0,\hdots,N$} and \mbox{$i = 1,\hdots,I$}, and combine them for all $i$ and a specific time step $k$ in the vectors $\lambda_k = (\lambda_{0,k},\hdots,\lambda_{I,k})^{\top} \allowbreak \in \mathbb{R}^{I}$, as well as further summarizing them in \mbox{$\lambda = (\lambda_{0}^{\top}, \hdots, \lambda_{N}^{\top})^{\top} \in  \mathbb{R}^{IN}$}. Additionally, we express the inequality constraints via \mbox{$G(x_{k},u_{k}) = (g_{0}(x_k,u_k), \hdots, g_{I}(x_k,u_k))^{\top} \in \mathbb{R}^{I}$}, such that the Lagrangian of problem \eqref{eq:optimizationproblem}   follows as
\begin{equation*}
    \mathcal{L}(\theta,\lambda,\mathcal{F}_{U,x_0},U) =  {\textstyle \sum_{k = 0}^{N-1}} (\theta^{\top} \phi(F_k(U,x_0),u_k) +
    \lambda_k^{\top}  
    G(F_k(U,x_0),u_k)).
\end{equation*}
Then, for all $k = 0,\hdots,N$ and $i = 1,\hdots,I$, the KKT conditions in accordance to the optimal demonstration $U^{*}$ and the initial condition $x_0^*$ are given as
\begin{subequations}
\begin{align}
    &\nabla_{U}\mathcal{L}(\theta,\lambda,\mathcal{F}_{U,x_0},U)\vert_{x_0 = x_0^*,U = U^*} = \mathbb{0}_{mN,1} \label{eq:kkt1u}\\
    & \lambda_{i,k} g_i(F_{k}(U^*,x_{0}^*),u_k^*) = 0 \label{eq:kkt2u}\\ 
    & g_i(F_{k}(U^*,x_{0}^*),u_k^*) \leq 0, \lambda_{i,k} \geq 0. \label{eq:kkt4u}
\end{align}
\label{eq:kkt}
\end{subequations}
\noindent Solving \eqref{eq:kkt} for $\theta$ and $\lambda$ returns the coefficient vector $\theta^*$ of the corresponding forward problem, as well as its Lagrangian multipliers $\lambda^*$.

\subsection{Sub-Optimal Demonstrations}
\label{subsec:suboptimaldemonstrations}
The considered sub-optimal demonstrations follow a unified structure given in Assumption~\ref{assumption:structuresuboptimal}.
\begin{assumption}
The available sequences, indicated with $U_{d}= (u_{d,0}^{\top}, \hdots, u_{d,N-1}^{\top})^{\top}$ for $d = 1,\hdots,D$, are noisy realizations of the optimal sequence $U^*$: \vspace{-0.4em}
\begin{align}
     U_{d} = U^* + n_{d}, \label{eq:noisyU}
\end{align}
where $n_d=(n_{d,0}, \hdots, n_{d,N-1})^{\top} \allowbreak \in \mathbb{R}^{Nm}$ is a vector collecting the realization of the additive noise.
\label{assumption:structuresuboptimal}
\end{assumption}
\vspace{-0.7em}
To address various causes of sub-optimal data, such as, e.g., a non-expert demonstrator or measurement noise, we consider two scenarios.
\paragraph*{Problem 1}
Demonstrations are i.i.d., normally distributed around $U^*$, i.e., $n_{d,k} \sim \mathcal{N}(\mathbb{0}_{m,1},\Sigma_{u})$ for all $d=1,...,D$ and $k=0,...,N-1$.  The covariance matrix $\Sigma_{u}$ thereby reflects the difficulty of obtaining an optimal input and is potentially unknown.
\paragraph*{Problem 2} Additive noises $n_{d,k}$,  for all $d=1,...,D$ and $k=0,...,N-1$, follow an unknown distribution.

\subsection{A Least Squares Approximation of Inverse KKT}
\label{subsec:aleastsquareapproximationofinversekkt}
Working with sub-optimal demonstrations not only requires a nontrivial constraint handling, but can additionally cause violations of the stationarity condition in equation \eqref{eq:kkt1u}, yielding
\begin{equation*}
    \nabla_{U}\mathcal{L}(\theta^{*}, \lambda^{*},\mathcal{F}_{U,x_0},U)\vert_{x_0 = x_{d,0},U = U_{d}} \neq \mathbb{0}_{mN,1}. 
    \label{eq:kkt1violated}
\end{equation*}
To deal with this issue, approaches such as those in \cite{englert2017,menner2019} propose to reformulate \eqref{eq:kkt1u} in terms of a least-squares optimization problem, which we indicate with $(\theta^{\top}, \lambda_{[1:D]}^{\top})^{\top} = KKT(U_{[1:D]})$. Its objective then reads~as
\vspace{-0.5em}
\begin{align}
        \min_{\theta,\lambda_{[1:D]}} \sum_{d=0}^D &\Vert\nabla_{U}\mathcal{L}(\theta,\lambda^{d},\mathcal{F}_{U,x_0},U)\vert_{x_0 = x_{d,0},U = U_{d}} \Vert^2 \label{eq:kktopt}
\end{align}
and conditions \eqref{eq:kkt2u} and \eqref{eq:kkt4u} are considered in the constraints.  However,  as detailed in Subsection~\ref{subsec:stationarityconditionasaneivregression}, the resulting problem has an EIV nature, and a plain least-squares estimate is not consistent -- i.e.,  it is biased and does not converge to its optimal value as $D \rightarrow +\infty$.  

\section{IOC as an Errors-in-Variables Problem}
In this section, we introduce the EIV-regression problem emerging from the inverse KKT problem with sub-optimal demonstrations, and present an IOC framework that takes inspiration from existing EIV solution strategies. The first approach builds upon an MCMC sampler, addresses Problem~1, and is presented in Section~\ref{subsec:mlwithcovupdate}. The second approach, presented in Section~\ref{subsec:tlswithexactpenalty}, addresses Problem 2 and follows the idea of total least squares.  Finally, practical aspects of the proposed framework are discussed in Section~\ref{subsec:practivalaspects}

\subsection{Stationarity Condition as an EIV-Regression}
\label{subsec:stationarityconditionasaneivregression} 
Denoting with $ J_{\theta} = \nabla_{U}(\sum_{k=0}^{N-1}\phi(F_k(U,x_0),u_k))$ and with $J_{\lambda} = \nabla_{U}(\sum_{k=0}^{N-1}G(F_k(U,x_0),u_k)))$,
the stationarity condition in \eqref{eq:kkt1u} can be interpreted as the following linear regression problem
\begin{equation}
    \underbrace{\begin{pmatrix}
    J_{\theta}  &  J_{\lambda}
    \end{pmatrix}\vert_{x_0 = x_0^*,\,U = U^{*}}
    }_{J(U)}
    \underbrace{
    \begin{pmatrix}
    \theta  \\ 
    \lambda
    \end{pmatrix}}_{\beta} = \underbrace{\mathbb{0}_{mN,1}}_{Y}.
    \label{eq:linearsystemofequationwrtu}
\end{equation}
There, $\beta$ acts as the regression parameter vector,  $J(U)$ as a stack of feature vectors, and $U$ as the independent variable vector. The dependent variables, collected in $Y$, are equal to zero when $U = U^*$.  However,  when operating with $U_d$ as in \eqref{eq:noisyU}, noise enters through the independent variable in the regressors, yielding an EIV 
problem. As a direct consequence, a naive least-squares estimate results biased and not consistent \citep{soderstrom2019}.  To address this issue, we present two optimization-based approaches in which the objective reflects the quality of a potential solution vector $\beta$ in accordance with \eqref{eq:linearsystemofequationwrtu} and the available data, and consider conditions \eqref{eq:kkt2u} and \eqref{eq:kkt4u} in the constraints.

\subsection{Approach 1: Maximum A Posteriori Estimate}
\label{subsec:mlwithcovupdate}
\begin{figure}[h]
\centering
\begin{tikzpicture}[->,>=stealth',shorten >=0.pt,auto,node distance=2cm,>=stealth',bend angle=45,auto]
\tikzstyle{place}=[circle,very thick,draw=black!75,fill=blue!20,minimum width=1cm]
\tikzstyle{placel}=[ellipse,very thick,draw=black!75,fill=blue!20]
\tikzstyle{red place}=[place,fill=red!20]
\tikzstyle{green place}=[place,fill=yellow!20]
\tikzstyle{transition}=[rectangle,very thick,draw=green!75,fill=blue!20,minimum size=2mm]
\tikzstyle{every label}=[red]
\begin{scope}[scale=0.75,transform shape]
\node [red place] (beta) {$\beta$};
\node [green place] (Yd) [right of=beta] {$\{Y_d\}_{d=1}^D$};
\node [red place] (U) [right of = Yd] {$U$};
\node [green place] (Ud) [right of = U] {$\{U_d \}_{d=1}^D$};
\node [red place] (SigmaU) [right of = Ud] {$\Sigma_U$};

\path (beta) edge node {} (Yd);
\path (U) edge node {} (Yd);
\path(U) edge node {} (Ud);
\path(SigmaU) edge node {} (Ud);
\end{scope}
\end{tikzpicture}
\caption{Visualization of the Bayesian network considered.} 
\label{fig:bayesnet}
\end{figure}
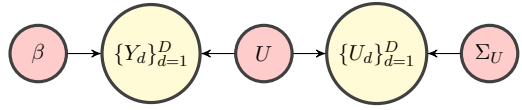

This first approach addresses Problem 1 and relies on the information about the distribution of the additive noise in \eqref{eq:noisyU}. The goal of jointly estimating $U^*$ and $\beta$ is formulated as a Maximum-a-Posteriori problem as follows.  \\We treat the independent variable $U$ as a random vector with prior distribution $U \sim \mathcal{N}(U_0, \Sigma_{U_0})$. It should capture the optimal $U^*$ and yield a likelihood model $U_d| U \sim \mathcal{N}(U, \Sigma_U)$ describing the suboptimal demonstrations according to \eqref{eq:noisyU}.  Moreover,  the likelihood associated with~\eqref{eq:linearsystemofequationwrtu} is $Y_d|U,\beta \sim \mathcal{N}(J(U)\beta, \Sigma_Y)$, where the covariance $\Sigma_Y$ describes the discrepancy from the vector $\mathbb{0}_{mN,1}$ and is deterministically chosen. 
To complete the Bayesian description of our problem, we consider the following priors: $\beta \sim \mathcal{N}(\beta_0, \Sigma_{\beta})$ and $\Sigma_U \sim IW(W_U, m_U)$. The resulting Bayesian network is depicted in Figure \ref{fig:bayesnet}. Noting that $U_d$ and $Y_d$ are conditionally independent given $U$, the posterior for the unknown $\beta$, $U$ and $\Sigma_U$ reads as follows:
\begin{align}
    & \;p(\beta,\Sigma_{U},U \vert \Sigma_{Y},Y_{[1:D]},U_{[1:D]}) \label{eq:posteriorprob}  \\ &\propto {\textstyle \prod_{d=1}^{D}} \mathcal{N}(Y_{d}; J(U)\beta, \Sigma_{Y})\cdot  \mathcal{N}(U_{d}; U, \Sigma_U)\cdot \mathcal{N}(U; U_{0}, \Sigma_{U_0})\cdot  \mathcal{N}(\beta; \beta_{0}, \Sigma_{\beta})\cdot IW(\Sigma_U; W_U, m_U). \nonumber
\end{align}
\vspace{-1.8em}

\noindent The rationale of our proposed approach consists in finding $U$ and $\beta$ maximizing such a posterior, while taking into consideration the constraints \eqref{eq:kkt2u} and \eqref{eq:kkt4u}. This Maximum-A-Posteriori approach can be written as follows:
\vspace{-0.7em}
\begin{equation}
    \min_{\beta,U}  \; P(U, \beta|\Sigma_U) \qquad \text{subject to } \begin{cases}
        &\lambda_{i,k} g_i(F_k(U,x_0),u_{k}) = 0 \\
        & g_i(F_k(U,x_0),u_{k}) \leq 0, \quad \lambda_{i,k} \geq 0  \\
        & k = 0,\hdots,N, i = 1,\hdots,I, 
    \end{cases}
    \label{eq:kkteiv}
\end{equation}
where the cost $P(U,\beta | \Sigma_U)$ is obtained by taking the negative logarithm of \eqref{eq:posteriorprob} and neglecting the distribution of $\Sigma_U$. 
Before solving \eqref{eq:kkteiv}, we perform a full Bayesian treatment of the problem to provide (A) an estimate for $\Sigma_U$ and (B) good candidate initial values for $U$ and $\beta$. This is done with an MCMC strategy based on a single-component Metropolis-Hastings sampling scheme~\citep{gilks1995}. We sketch the procedure in the following paragraph, and defer to the Supplementary material for the detailed derivation.

\paragraph*{MCMC strategy}
The (single-component) Metropolis-Hastings algorithm allows for the generation of a Markov Chain for which the invariant distribution is a target distribution of interest. In our case, we aim at simulating the posterior \eqref{eq:posteriorprob}. This is done by iteratively sampling from the so-called  \textit{full conditionals}, which in this case are $p(\beta | Y_{[1:D]}, \Sigma_Y)$, $p(U|U_{[1:D]},Y_{[1:D]},\beta, \Sigma_U)$ and $p(\Sigma_U|U_{[1:D]},U)$.
We first focus on the case in which the dynamics in \eqref{eq:dynamics} are linear, the objective features in $\phi(\cdot)$ are quadratic, and constraints in $\{ g_i(\cdot,\cdot)\}_{i=1}^I$ are polytopic. In this scenario, $J(U)$ can be re-written as an affine transformation of $U$, i.e., $J(U)\beta = (MU + E)\beta$. The full conditionals, inspecting the likelihood-prior products, read as follows:
\begin{equation}
p(\beta|Y_{[1:D]},U) \propto  \prod_{d=1}^D \mathcal{N}\Big(Y_d;(MU + E)\beta, \Sigma_Y\Big) \cdot \mathcal{N}(\beta;\beta_0,\Sigma_{\beta}) \label{eq:fullbeta}
\end{equation}
\vspace{-0.5em}
\begin{equation}
p(U|\beta, Y_{[1:D]},U_{[1:D]},\Sigma_U) \propto \prod_{d=1}^D \mathcal{N}(Y_d;(MU + E)\beta, \Sigma_Y)\cdot \mathcal{N}(U_d;U,\Sigma_U)\cdot \mathcal{N}(U; U_0, \Sigma_{U_0}) 
\label{eq:fullU}
\end{equation}
\vspace{-0.5em}
\begin{equation}
p(\Sigma_U|U_{[1:D]}, U) \propto \prod_{d=1}^D \mathcal{N}(U_d; U, \Sigma_U)\cdot IW(\Sigma_U; W_U, m_U).\label{eq:fullSigmaU}
\end{equation}
By leveraging standard properties of conjugate priors \citep{gelman2004}, the distributions above have a closed-form expression as Gaussian (\eqref{eq:fullbeta}, \eqref{eq:fullU}) and Inverse Wishart \eqref{eq:fullSigmaU}. Since these are easy to sample, the single-component Metropolis-Hastings scheme becomes a Gibbs sampler. Because all the involved distributions are well-defined, the stochastic simulation scheme is ergodic; at convergence,
we perform a Monte Carlo integration of the full conditionals to obtain the sample mean for $\Sigma_U$, $\beta$ and $U$. The first is used to set the cost in \eqref{eq:kkteiv}, while the other two are used as intialization of the solver. With this construction, the following Theorem holds. Its proof can be found in the Appendix.
\begin{theorem}
Consider linear system dynamics, polytopic constraints, and quadratic features in $\phi$. Then, if the adopted solver for problem \eqref{eq:kkteiv} converges to the global minimum, its solution returns the maximum a posteriori estimate of the posterior probability function in \eqref{eq:posteriorprob}, fulfilling conditions \eqref{eq:kkt2u} and \eqref{eq:kkt4u}. Furthermore, the estimate is consistent. \label{theorem:convergence}
\end{theorem}
In the more general case of nonlinear dynamics, \eqref{eq:fullU} is not Gaussian anymore, so a Metropolis-Hastings step has to be adopted. As a further consequence, the cost loses bi-convexity, and convergence is ensured only to a local optimum.\\
Finally, note that the sampling-based solution can easily be extended to non-Gaussian (but known) distributions by applying a general Metropolis-Hastings step on the required full-conditionals.

\subsection{Approach 2: Total Least Square Estimate}
\label{subsec:tlswithexactpenalty}
This section addresses Problem 2, where a likelihood function cannot be constructed because no information on the noise distribution is available. In this approach, we employ total least squares to estimate the regression parameter vector $\beta$ and the residuals $r_{[1:D]}$, which describe the noise realizations. A first formulation reads as follows:
\begin{equation}
    \min_{r_{[1:D]},\beta} \sum_{d =1}^{D} r_d^{\top}\Sigma_{U}^{-1}r_d \quad \text{subject to } \begin{cases}
        J(U_{d} - r_d)\beta=\mathbb{0}_{mN,1}  \\
    \lambda_{i,k} g_i(F_k(U_{d}-r_{d},x_0),u_{d,k}-r_{d,k}) = 0 \\ 
    g_i(F_k(U-n_{d},x_0),u_{d,k}-r_{d,k}) \leq 0, \lambda_{i,k} \geq 0 \\
k = 0,\hdots,N, i = 1,\hdots,I, d = 1,\hdots,D.
    \end{cases}
\label{eq:kktlsresiduals}
\end{equation}
To reduce the number of necessary optimization variables, we leverage the assumption that a unique optimal demonstration $U^{*}$ exists, and extend the optimization problem in \eqref{eq:kktlsresiduals} with the equality constraint $U = U_{d} -r_{d}$ for all $d=1,...,D$. This allows for the transformation of \eqref{eq:kktlsresiduals} into an equivalent problem, optimizing for a single demonstration $U$ instead of all residuals by replacing each usage of $U_{d}-r_{d}$ with $U$. Thus, \eqref{eq:kktlsresiduals} becomes
\begin{equation}
    \min_{U,\beta} \sum_{d =1}^{D} (U-U_{d})^{\top}\Sigma_{U}^{-1}(U-U_{d}) \quad \text{subject to } \begin{cases} J(U)\beta=\mathbb{0}_{mN,1} \\
    \lambda_{i,k} g_i(F_k(U,x_0),u_{k}) = 0 \\ 
g_i(F_k(U,x_0),u_{k}) \leq 0, \lambda_{i,k} \geq 0  \\
    k = 0,\hdots,N, i = 1,\hdots,I.
    \end{cases}
    \label{eq:kktlspre}
\end{equation}
However, such a problem depends on $\Sigma_U$, which is initially unknown. To obviate this issue, we propose an alternating scheme which (A) starts from an initial guess for $U$ and $\beta$, (B) estimates the unknown covariance as $\Sigma_U = \frac{1}{D}{\textstyle \sum_{d = 1}^{D}} (U-U_{d})^{\top}(U-U_{d})$, (C) solves \eqref{eq:kktlspre}, and (D) iterates the steps (B-C). The performance of such an approach is summarized in the following Theorem, which is proved in the Appendix. 
\begin{theorem}
    Consider linear system dynamics, polytopic constraints, and quadratic features in $\phi$. Furthermore, assume that $n_k$ has mean zero. Then, if the solver for problem \eqref{eq:kktlspre} converges to the global minimum, the proposed procedure returns a consistent estimate for $\beta$.
\label{theorem:convergence2}
\end{theorem}

\subsection{Practical Aspects}
\label{subsec:practivalaspects}
In this section, we discuss two main considerations arising in the developed optimization schemes: namely, the initialization and the adopted solvers.

\subsubsection{Initial Guess} 
\label{subsubsec:initialguess}
Especially for non-convex optimization problems, or problems with only a few available demonstrations, a suitable initial guess and solver initialization are critical to improve the final estimate or reduce computation time.  
If no prior knowledge is available, we suggest setting $U_0$ equal to the sample mean, 
and $\beta_0$ as its corresponding estimate $KKT(U_0)$. The matrices $\Sigma_Y$ and $W_U$ can be either set to a covariance approximation with respect to $U_{0}$, risking exaggerated certainty of the estimated parameters \citep{kass1989}, or to a diagonal matrix, reflecting the independence of the investigated noise. 
\subsubsection{Solver}
\label{subsection:handelingofeuqilibriumconstraints}
For linear demonstrating system dynamics, polytopic constraints and quadratic features in $\phi(\cdot)$, the optimization problems in \eqref{eq:kkteiv} and \eqref{eq:kktlspre} are bi-convex and fulfill all requirements for global convergence of the Global Optimization Algorithm \citep{floudas2000}. In any other case, they denote a general Mathematical Problem with Equilibrium Constraints (MPEC). While such problems can be hard for most common solvers (e.g., IPOPT \citep{wachter2006} or SQP-based ones \citep{boggs1995}), the strongest convergence guarantees are provided for combinatorial methods, such as pivoting \citep{fang2012} or active set methods \citep{giallombardo2008,leyffer2007}. The iterative application of nonlinear programming methods to a relaxed version of the MPECs is actively investigated \citep{scholtes2001,kadrani2009,kanzow2013}, leading to the often-used IPOPT-C solver \citep{raghunathan2005}. In this work, the MPEC is addressed by combining the relaxation approach by \cite{scholtes2001} with IPOPT. The relaxation constant and its decrease factor entering the formulation are individually adjusted for each experiment, which is a key and delicate task to ensure that the solver converges to an acceptable solution.  
\section{Experimental Results}
The proposed framework is tested on three different systems in simulation: a spring-damper system (Section \ref{sec:springdamper}), the kinematic bicycle model (Section \ref{sec:bycicle}), and a two-compartment Bayesian glucose model (Section \ref{sec:glucose}). The continuous system dynamics are discretized using the backward Euler method with sampling time \mbox{$T_s = 0.1$} for the first two, and \mbox{$T_s = 1$} for the third system. While for the first two systems we employ Approach 1 (Section \ref{subsec:mlwithcovupdate}), in the third we use Approach 2 (Section \ref{subsec:tlswithexactpenalty}) to respect the strictly positive insulin input that results in a non-Gaussian input noise. Throughout the experiments, we consider different values for the input noise covariance, which are calculated with respect to the mean input values of the optimal demonstration over the considered horizon; we denote such a mean value with $u_{m,D}$.
All estimates obtained with our framework are further denoted with  $U_{\mathrm{EIV}}$ and  $\theta_{\mathrm{EIV}}$, and their quality is evaluated by means of the root mean square error (RMSE) with respect to the true values $U^{*}$ and $\beta^{*}$. In each set-up we repeat the experiments 10 times, and compare the performance against the $\beta$ obtained with the inverse KKT least square relaxation proposed in \cite{menner2019}, and the mean of all demonstrations $U_{m} = \frac{1}{D}\sum_{d=1}^{D}U_{d}$. 

\subsection{Spring-Damper System}\label{sec:springdamper}
\noindent The system dynamics read as 
\begin{align}
    & \dot{x}_1 = x_2 ,\ \ \ \dot{x}_2 = \frac{1}{m}(-cx_1 -dx_2 + u), \nonumber
\end{align}
with $m=1.0$ $\mathrm{kg}$, $c=0.2$ $\mathrm{kg\cdot s^{-2}}$ and $d=0.1$ $\mathrm{kg\cdot s^{-1}}$.
The optimal control forward problem has a horizon length of $N = 10$ and includes the inequality constraint $u \leq 0.7$. The unknown parameter is set to $\theta = (10,5,7)^{\top}$, the feature vector to $\phi = ((x_1-3)^{2},(x_2-0)^{2})^{\top}$, and the initial value is chosen as $x_0 = (1,0.1)^{\top}$. Mean and standard deviations of the RMSE are presented in Table \ref{tab:springdamp}.\\
\vspace{-1em}
\begin{table}[h]
\small
\begin{center}
\begin{adjustbox}{width=0.65\textwidth}
\begin{tabular}{|C{0.7cm}|C{0.8cm}|C{2.9cm}|C{2.9cm}|C{2.9cm}|C{2.9cm}|}
\hline
\multicolumn{2}{|c|}{Method }  & \multicolumn{3}{c|}{$\Sigma_{U}$} \\
\cline{3-5}
\multicolumn{2}{|c|}{}                   & 5\% $u_{m,D}$ & 10\% $u_{m,D}$  & 20\% $u_{m,D}$ \\
\hline
\rowcolor[HTML]{d8eddf} 
\cellcolor[HTML]{FFFFFF}
&$\theta_{\mathrm{KKT}}$ & $2.38 \pm 0.991$ & $1.41 \pm 0.859$ & $2.89 \pm 2.009$ \\
\cline{2-5}
\rowcolor[HTML]{EFEFEF} 
\multirow{-2}{*}{\cellcolor[HTML]{FFFFFF}KKT} &$U_{\mathrm{m}}$ & $0.024 \pm 0.005$ & $0.05 \pm 0.009$ & $0.11 \pm 0.028$ \\
\hline
\rowcolor[HTML]{d8eddf} 
\cellcolor[HTML]{FFFFFF}
& $\boldsymbol{\theta_{\mathrm{EIV}}}$  & $\mathbf{0.57 \pm 0.374}$ & $\mathbf{0.73 \pm 0.642}$ & $\mathbf{1.05 \pm 0.488}$  \\ \cline{2-5} 
\rowcolor[HTML]{EFEFEF} 
\multirow{-2}{*}{\cellcolor[HTML]{FFFFFF}\textbf{EIV}} &$\mathbf{U}_{\mathrm{\textbf{EIV}}}$ & $\mathbf{0.01 \pm 0.005}$ & $\mathbf{0.02 \pm 0.008}$ & $\mathbf{0.04 \pm 0.027}$ \\ \hline
\end{tabular}
\end{adjustbox}
\caption{Mean and standard deviations of the RMSE for parameter and trajectory estimates on the spring-damper system.}
\end{center}
\label{tab:springdamp}
\end{table}
\normalsize

\vspace{-2em}
\subsection{Kinematic Bicycle Model}\label{sec:bycicle}
\noindent Representing a more realistic use case, Approach 1 has been tested with the kinematic bicycle model \citep{Rajamani2012}. The demonstrating system dynamics read as:
\begin{align}
    \dot{x}_1 = u_1 \cos(x_3), \ \ \ \dot{x}_2 = u_1 \sin(x_3), \ \ \  
    \dot{x}_3 = {u_1} \tan(x_4)/L, \ \ \
    \dot{x}_4 = u_2. \nonumber
 \end{align}
We set $L = 0.115$ $\mathrm{m}$. The considered forward problem has a horizon length of $N = 10$ and includes the inequality constraint $u_1 \leq 2.2$. The unknown parameter is set to $\theta = (10,10,3,3,8,5)$ and the input noise acts on the steering input $u_2$. The feature vector is set to $\phi = ((x_1-3)^{2},(x_2-3)^{2},(x_3-0)^{2},(x_4-0)^{2})^{\top}$ and the initial value is chosen as $x_0 = (0,0,0,0)^{\top}$. Mean and standard deviations of the RMSE are presented in Table~\ref{tab:kinematicbicycle}.

\begin{table}[h]
\small
\begin{center}
\begin{adjustbox}{width=0.65\textwidth}
\begin{tabular}{|C{0.7cm}|C{0.8cm}|C{2.9cm}|C{2.9cm}|C{2.9cm}|C{2.9cm}|}
\hline
\multicolumn{2}{|c|}{Method}  & \multicolumn{3}{c|}{$\Sigma_{U}$} \\
\cline{3-5}
\multicolumn{2}{|c|}{}                   & 5\% $u_{m,D}$ & 10\% $u_{m,D}$ & 20\%  $u_{m,D}$\\
\hline
\rowcolor[HTML]{d8eddf} 
\cellcolor[HTML]{FFFFFF}
 &$\theta_{\mathrm{KKT}}$ & $3.20 \pm 0.001$ & $3.20 \pm 0.002$ & $3.18 \pm 0.007$ \\
\cline{2-5} 
\rowcolor[HTML]{EFEFEF} 
\multirow{-2}{*}{\cellcolor[HTML]{FFFFFF}KKT}
&$U_{\mathrm{m}}$ & $0.0006 \pm 0.0001$ & $0.001 \pm 0.0002$ & $0.002 \pm 0.0005$ \\
\hline
\rowcolor[HTML]{d8eddf} 
\cellcolor[HTML]{FFFFFF}  &$\boldsymbol{\theta_{\mathrm{EIV}}}$ & $\mathbf{0.29 \pm 0.096}$ & $\mathbf{0.48 \pm 0.161}$ & $\mathbf{0.98 \pm 0.114}$ \\
\cline{2-5}
\rowcolor[HTML]{EFEFEF} 
\multirow{-2}{*}{\cellcolor[HTML]{FFFFFF}\textbf{EIV}}
&$\mathbf{U}_{\mathrm{\textbf{EIV}}}$ & $\mathbf{0.0007 \pm 0.0001}$ & $\mathbf{0.001 \pm 0.0002}$ & $\mathbf{0.002 \pm 0.0004}$ \\
\hline
\end{tabular}
\end{adjustbox}

\caption{Mean and standard deviations of the RMSE for parameter and trajectory estimates on the kinematic bicycle model.}
\label{tab:kinematicbicycle}
\end{center}
\end{table}
\normalsize

\subsection{Two-Compartment Bayesian Glucose Model}\label{sec:glucose}
\noindent As biomedical applications are considered an important potential use case of inverse optimal control, we finally test the approach proposed in Section \ref{subsec:tlswithexactpenalty} on the two-compartment Bayesian glucose model from \cite{callegari2003}. The demonstrating system dynamics are
 \begin{align}
    \dot{x}_1 = (-p_1 - k_{21} - x_3)x_1 + k_{12}x_2 + p_1Q_{1b}, \ \ \ \dot{x}_2 = k_{21}x_1 - k_{12}x_2, \ \ \ \dot{x}_3 = -p_2x_3 + p_3(u - I_b). \nonumber
 \end{align}
 For a full explanation on these equations, we defer to \cite{callegari2003}, and 
the parameter values are set according to patient 1 in Table 1 reported therein.
The forward problem has a horizon length of $N = 20$ and the insulin input is constrained as $u \geq 0$. The unknown parameter is set to $\theta = (1,0.1,10)$, the feature vector to $\phi = ((x_1-Q_{1b})^{2},(x_2-\frac{k_{21}}{k_{12}}Q_{1b})^{2})^{\top}$ and the initial value is chosen as $x_0 = (Q_{1b} + 330,\frac{k_{21}}{k_{12}}Q_{1b})^{\top}$. Mean and standard deviations of the RMSE are presented in Table \ref{tab:glucose}. 
Due to the nonlinearity in the dynamics, this model required a very careful tuning of the MPEC relaxation parameters to ensure convergence to a suitable stationary point. We argue that such a difficulty could be overcome by using a different solver.

\begin{table}[h]
\small
\begin{center}
\begin{adjustbox}{width=0.65\textwidth}
\begin{tabular}{|C{0.7cm}|C{0.8cm}|C{2.9cm}|C{2.9cm}|C{2.9cm}|C{2.9cm}|}
\hline
\multicolumn{2}{|c|}{Method}  & \multicolumn{3}{c|}{$\Sigma_{U}$} \\
\cline{3-5}
\multicolumn{2}{|c|}{}                   & 5\% $u_{m,D}$ & 10\% $u_{m,D}$ & 20\% $u_{m,D}$ \\
\hline
\rowcolor[HTML]{d8eddf} 
\cellcolor[HTML]{FFFFFF}
&$\theta_{\mathrm{KKT}}$ & $0.12 \pm 0.007$ & $0.13 \pm 0.015$ & $0.24 \pm 0.046$ \\
\cline{2-5}
\rowcolor[HTML]{EFEFEF} 
\multirow{-2}{*}{\cellcolor[HTML]{FFFFFF}KKT}
&$U_{\mathrm{m}}$ & $0.70 \pm 0.091$ & $1.40 \pm 0.123$ & $3.25 \pm 0.182$ \\
\hline
\rowcolor[HTML]{d8eddf} 
\cellcolor[HTML]{FFFFFF}  &$\boldsymbol{\theta_{\mathrm{EIV}}}$ & $\mathbf{0.09 \pm 0.042}$ & $\mathbf{0.10 \pm 0.033}$ & $\mathbf{0.12 \pm 0.015}$ \\
\cline{2-5}
\rowcolor[HTML]{EFEFEF} 
\multirow{-2}{*}{\cellcolor[HTML]{FFFFFF}\textbf{EIV}} 
&$\mathbf{U}_{\mathrm{\textbf{EIV}}}$ & $\mathbf{0.10 \pm 0.061}$ & $\mathbf{0.20 \pm 0.113}$ & $\mathbf{0.41 \pm 0.320}$ \\
\hline
\end{tabular}
\end{adjustbox}
\vspace{0.5em}
\caption{Mean and standard deviations of the RMSE for parameter and trajectory estimates for the two-compartment glucose model of \cite{callegari2003}.}
\end{center}
\label{tab:glucose}
\end{table}
\normalsize
\vspace{-3em}

\section{CONCLUSIONS}

We presented two IOC approaches for the estimation of an unknown parameter vector in partially known objectives from sub-optimal demonstrations. Both strategies build upon the KKT conditions and allow for the consideration of inequality constraints in the optimization problem of interest. The key idea in the proposed methods consists in addressing the EIV nature of the problem to obtain unbiased estimates. We consider two scenarios: in the first, we assume that the input noise entering the dynamics is distributed according to a Gaussian (but it could be extended to any known distribution), while in the second such information is unavailable. We tackle the first situation with a Bayesian strategy, leveraging MCMC to initialize the actual constrained optimization problem, while for the second we deploy a formulation based on total least squares. Differently from other approaches in the literature, both of our proposed strategies learn the noise input covariance from data; additionally, the first approach can also include prior information and allow for uncertainty quantification. Theoretical consistency guarantees are provided for linear systems with zero-mean additive noise, polytopic constraints, and quadratic objectives. Results in simulation 
show that (i) the estimated input sequence is closer to the optimal one with respect to a naive sample mean of the demonstrating sequences, and (ii) the proposed approaches outperform a previously presented method relying on a least-squares relaxation of the classical KKT inversion approach, especially in scenarios with higher input noise levels.

\section*{Appendix}

In the following, we present the proofs of Theorems \ref{theorem:convergence} and~\ref{theorem:convergence2} stated in Sections \ref{subsec:mlwithcovupdate} and \ref{subsec:tlswithexactpenalty}, respectively. 
\subsection*{Proof of Theorem \ref{theorem:convergence}}
Given the irreducibility and aperiodicity of the chosen MCMC sampling algorithms, obtaining a maximum a posteriori estimate follows directly from the ergodic theorem, as well as global convergence of the bi-convex problem in \eqref{eq:kkteiv}. To prove the consistency of this estimate, we reformulate the objective in \eqref{eq:kkteiv} with respect to an arbitrary positive definite covariance and divide it by the amount of data $D$. This reads as
{\small
\begin{align*}
     -\Bigg(&\sum_{d = 1}^{D} \frac{1}{D} \Big((J(U)\beta)^{\top}\Sigma_{Y}^{-1}(J(U)\beta) + Y_{d}^{\top}\Sigma_{Y}^{-1}Y_{d} - 2(J(U)\beta)^{\top}\Sigma_{Y}^{-1}Y_{d}   + U^{\top}\Sigma_{U}^{-1}U \\  &- 2U^{\top}\Sigma_{U}^{-1}U_{d}  +  U_{d}^{\top}\Sigma_{U}^{-1} U_{d}\Big)  + \frac{1}{D}((U - U_{0})^{\top} \Sigma_{U_0}^{-1} (U - U_{0}) + (\beta -\beta_{0})^{\top} \Sigma_{\beta_{0}}^{-1} (\beta - \beta_{0}))\Bigg).\nonumber 
\end{align*}
}%
By the Law of Large Numbers, acknowledging that the mean of $Y$ is equal to $0$, and replacing $U_{d}$ with $U^{*} + N_{U,d}$, then for $D \rightarrow \infty$ we obtain
\vspace{-0.5em}
\begin{align}
    &(J(U)\beta)^{\top}\Sigma_{Y}^{-1}(J(U)\beta) + 
U^{\top}\Sigma_{U}^{-1}U - 2U^{\top}\Sigma_{U}^{-1}U^{*}  + U^{*\top}\Sigma_{U}^{-1} U^{*} + C \nonumber \\
=& \ (J(U)\beta)^{\top}\Sigma_{Y}^{-1}(J(U)\beta) + (U - U^{*})^{\top}\Sigma_{U}^{-1}(U - U^{*}) + C,  \label{eq:lastcost}
\end{align}
where $C$ is a constant term that depends on $\beta$ and $U$. The proof is concluded by noting that \eqref{eq:lastcost} is minimized by taking $U^{*}$ and $\beta^{*}$. \hfill $\blacksquare$

\subsection*{Proof of Theorem \ref{theorem:convergence2}}
 The proof of Theorem \ref{theorem:convergence2} directly follows along the line of the consistency proof in Theorem~\ref{theorem:convergence}, showing that the cost function of optimization problem \eqref{eq:kktlspre} is minimized by $U^*$. Consequently, the estimate of $\beta$ follows as $\beta^*$. \hfill $\blacksquare$

\newpage

\section*{Supplementary material}
\appendix 
\noindent
In this document we detail the Markov Chain Monte Carlo (MCMC) procedure presented in Section \ref{subsec:mlwithcovupdate} of the main paper. 

\section{Synopsis of MCMC}\label{sec:MCMCprimer}
\paragraph*{Scope} The starting problem can be stated as follows. Given a random vector $Z \in \mathcal{Z} \subseteq \mathbb{R}^M$, distributed according to a known probability density $\pi$, we aim at computing 
\begin{equation}
    \mathbb{E}[g(Z)] = \int_{\mathcal{Z}} g(z)\pi(z)dz. \label{eq:targetMCMC}
\end{equation}
For instance, $\pi(Z)$ is some posterior distribution, and \eqref{eq:targetMCMC} with $g(z)=z$ returns the Minimum Mean Squared Error estimate for $Z$~\citep{anderson2005}.\\ 
\paragraph*{Monte Carlo approximation} Typically, however, such an integral is difficult to solve analytically. To overcome this issue, Monte Carlo integration can be deployed to obtain a sample-based approximation given by
\begin{equation}
    \frac{1}{N_S}\sum_{j=1}^{N_S} g(z(j)) \quad \text{with i.i.d. }  z(j) \sim \pi. \label{eq:MonteCarloInt}
\end{equation}
By virtue of the Law of Large Numbers, the expression in \eqref{eq:MonteCarloInt} converges to \eqref{eq:targetMCMC} as the number of samples $N_S$ goes to infinity \citep{robert2004}. 
\paragraph*{The issue of sampling}  
The sampling from $\pi$ presents two possible challenges: (a) the density $\pi$ might have an involved expression, so it is difficult to sample; (b) obtaining i.i.d.~samples might be difficult, especially when the vector dimension $M$ is high. The idea to address these issues consists in building a Markov Chain, whose invariant distribution\footnote{The basic theory can be reviewed, e.g., in Chapter 6 of \cite{robert2004}.} equals the one associated with~$\pi$: at steady-state, samples from such a stochastic process still allow the Law of Large Numbers to hold \citep[Chapter 1.3.2]{gilks1995}. With this mechanism, issue (b) of having i.i.d.~samples is relaxed; now we address point (a), which is typically tackled with the Metropolis-Hastings algorithm \citep{chib1995}.
\paragraph*{The Metropolis-Hastings Algorithm} This procedure was introduced in \cite{metropolis2004} and generalized in \cite{hastings1970} (see also \cite{tierney1994} for a full historical overview). It is an iterative procedure, whose core is an acceptance-rejection mechanism for samples drawn from a user-chosen \textit{proposal density} $q(\cdot)$ for $Z$. The overall procedure is summarized in Algorithm \ref{alg:genMH}.
Note that the proposal density $q(\cdot)$ can also be independent of the current sample: if this is the case, the strategy is known as an \textit{independence sampler} \citep[Chapter 1.4.1]{gilks1995}.

\paragraph*{Dealing with high-dimensional vectors: the single-component version} When the original vector is high-dimensional, it is often difficult to find a good proposal density $q(\cdot)$ yielding a suitable acceptance rate (which should empirically be around 40\%). To obviate this issue, one can consider (blocks of) components of $Z$ and apply the Metropolis-Hastings to each of them. Specifically, assume that $Z$ is divided in $C$ blocks of components in the following way: $Z = [Z_1^{\top},...,Z_C^{\top}]^{\top}$, with $Z_{\ell} \in \mathbb{R}^{M_{\ell}}$ for each $\ell=1,...,C$, with $\sum_{\ell=1}^C M_{\ell} = M$. Denote with $Z_{-\ell}$ the vector $Z$ without the $\ell$-th component. The idea is then to apply the Metropolis-Hastings to each \textit{full conditional}, which reads, for the $j-$th iteration, as 
\begin{equation}
    \pi(Z_{\ell}(j)| Z_1(j),...,Z_{\ell-1}(j), Z_{\ell+1}(j-1),...,Z_{C}(j-1)).\notag
\end{equation}
Overall, the strategy is presented in Algorithm \ref{alg:SCMH}.

\paragraph*{Simple full conditionals: the Gibbs sampler} Sometimes, the full conditionals admit closed-form expressions by means of simple distributions, such as the Gaussian, the Gamma, or the Inverse-Wishart. This usually happens when the prior-likelihood pair yielding the full conditional (that can indeed be interpreted as a posterior) is a conjugate pair \citep{gelman2013}. There are two cases that are of particular interest for the approach presented in Section \ref{subsec:mlwithcovupdate} of the main paper. The first is when the likelihood is given by a Gaussian with unknown mean and known covariance, and the unknown mean is given a Gaussian prior; the second is when the likelihood is Gaussian with known mean and unknown covariance matrix, and the latter is given an Inverse Wishart prior. The full derivation can be found in Chapters 3.5 and 3.6 in \cite{gelman2013}, but the main results are summarized in Table \ref{tab:conjpriors}.
\begin{table}[H]
\centering
\begin{adjustbox}{width=\textwidth}
\begin{tabular}{|c|c|c|}
\hline
       Likelihood & Prior & Posterior \\
     \hline
     $Y|X \sim \mathcal{N}(SX, V)$ & $X\sim \mathcal{N}(m,P)$ & $X|Y \sim \mathcal{N}\Big(m + (P^{-1} + S^{\top}VS)^{-1}S^{\top}V^{-1}(Y-Sm), \ (P^{-1} + S^{\top}VS)^{-1}\Big)$\\
     \hline
     $X|P \sim \mathcal{N}(m,P)$ & $P\sim IW(W,\nu)$ & $P|X \sim IW(W + (X-m)(X-m)^{\top}, n_X + \nu)$\\
     \hline
\end{tabular}
\end{adjustbox}
    \caption{Useful conjugate pairs. The notation follows the one introduced in the corresponding paragraph in the Introduction to the main paper. The dummy variables used in this table have dimensions $X \in \mathbb{R}^{n_X}$ and $Y \in \mathbb{R}^{n_Y}$.}
    \label{tab:conjpriors}
\end{table}
\noindent
As Gaussians and Inverse Wishart are easy to sample, we use them directly instead of a proposal density: as a result, the acceptance probability in Algorithm \ref{alg:SCMH}  becomes always equal to~1.

\paragraph*{Summary} Thus, the MCMC strategy used in the approach proposed in Section \ref{subsec:mlwithcovupdate} of the main paper can be summarized as follows: (1) deploy a Gibbs sampler (possibly with a Metropolis-Hastings step on some of the full conditionals) to obtain the Markov chain, having as invariant distribution the posterior of interest; (2) use Monte Carlo integration for the variables that have to be estimated.
 \newpage

\begin{algorithm}[H]
\SetAlgoLined
    \KwData{Proposal density $q(\cdot|\cdot)$, target density $\pi(\cdot)$.}
    \KwResult{Samples $\{z(j)\}$ from $\pi(\cdot)$.}

 Initialize $z(0)$ and set $j=0$\;

\While{not converged}{
Sample a point $\bar{z}$ from the proposal density $q(\cdot|z(j))$\;

Sample a variable $u\sim\mathcal{U}([0,1])$\;

Compute the acceptance probability $$\boldsymbol{\mathrm{p}}_{a} = \min\Bigg(1, \frac{\pi(\bar{z})q(z(j)|\bar{z})}{\pi(z(j))q(\bar{z}|{z(j)})} \Bigg)$$\;

\eIf{$u \leq \boldsymbol{\mathrm{p}}_{a}$}{
Set $z(j+1)=\bar{z}$\;} 
{Set $z(j+1) = z(j)$\;
}
 $j \leftarrow j+1$\;
}
\caption{Metropolis-Hastings algorithm to explore $\pi(Z)$}
\label{alg:genMH}
\end{algorithm}

\begin{algorithm}[H]
\SetAlgoLined
\KwData{Proposal densities $q_{\ell}(\cdot|\cdot)$, full conditionals $\pi(\cdot|Z_{-\ell})$, $\ell=1,...,C$.}
    \KwResult{samples $\{z(j)\}$ from $\pi(\cdot)$.}
Initialize $z(0)$ and set $j=0$\;

\While{not converged}{
\For{$\ell=1,...,C$}{
Sample a point $\bar{z}_{\ell}$ from the proposal density $q_{\ell}(\cdot|z_{\ell}(j))$\;

Sample a variable $u\sim \mathcal{U}([0,1])$\;

Compute the acceptance probability $$\boldsymbol{\mathrm{p}}_{a,\ell} = \min\Bigg(1, \frac{\pi(\bar{z}_{\ell}|Z_{-\ell})q(z_{\ell}(j)|\bar{z}_{\ell})}{\pi(z(j)|Z_{-\ell})q(\bar{z}_{\ell}|{z_{\ell}(j)})} \Bigg)$$\;

\eIf{$u \leq \boldsymbol{\mathrm{p}}_{a,\ell}$}{
Set $z_{\ell}(j+1)=\bar{z}_{\ell}$\;} 
{Set $z_{\ell}(j+1) = z_{\ell}(j)$\;
}
}
 $j \leftarrow j+1$\;
}
\caption{Single-component Metropolis-Hastings}
\label{alg:SCMH}    
\end{algorithm}

\newpage
\section{Details for the Maximum-A-Posteriori approach based on MCMC, Section \ref{subsec:mlwithcovupdate}}

\paragraph*{Overview}
The optimization problem to be solved is given in \eqref{eq:kkteiv}, and reads as follows:
\begin{equation}
    \min_{\beta,U}  \; P(U, \beta \ | \ \Sigma_U) \qquad \text{subject to } \begin{cases}
        &\lambda_{i,k} g_i(F_k(U,x_0),u_{k}) = 0 \\
        & g_i(F_k(U,x_0),u_{k}) \leq 0, \quad \lambda_{i,k} \geq 0  \\
        & k = 0,\hdots,N, i = 1,\hdots,I,
    \end{cases}
    \notag
\end{equation}
where the cost is derived from the posterior $p(\beta,\Sigma_{U},U \vert \Sigma_{Y},Y_{[1:D]},U_{[1:D]})$. The idea is to adopt a full Bayesian treatment of the unknown variables to estimate, and use the MCMC procedure reviewed in Section \ref{sec:MCMCprimer} to obtain (A) the estimate for $\Sigma_U$ entering the cost, and (B) an initialization for the value of $U$ to be used in the solver for \eqref{eq:kkteiv}. 

\paragraph{Hypothesis entering the Bayesian construction} We now detail the expression for the posterior $p(\beta,\Sigma_{U},U \vert \Sigma_{Y},Y_{[1:D]},U_{[1:D]})$.\\
The available likelihoods are the following:
\begin{itemize}
    \item $U_d|U,\Sigma_U \sim \mathcal{N}(U,\Sigma_{U})$, i.i.d.~for all $d=1,...,D$. This models the suboptimal demonstrations;
    \item $Y_d|U,\beta \sim \mathcal{N}(J(U)\beta, \Sigma_Y)$, where $\Sigma_Y$ is the covariance of the zero-mean additive Gaussian noise, artificially added to model the discrepancy from 0 in \eqref{eq:linearsystemofequationwrtu}.
\end{itemize}
The adopted priors for the unknown variables are chosen as:
\begin{itemize}
    \item $\Sigma_U \sim IW(W_U,m_U)$;
    \item $\beta \sim \mathcal{N}(\beta_0,\Sigma_{\beta})$;
    \item $U \sim \mathcal{N}(U_0, \Sigma_{U_0})$.
\end{itemize}
Writing the posterior $p(\beta,\Sigma_{U},U \vert \Sigma_{Y},Y_{[1:D]},U_{[1:D]})$ as proportional to its likelihood-prior products following the dependencies depicted in the Bayesian network in Figure \ref{fig:bayesnet}, one obtains \eqref{eq:posteriorprob}. 

\paragraph{Resulting expression for the cost in \eqref{eq:kkteiv}} If $\Sigma_U$ is known, then we have 
\begin{equation}
   p(\beta, U | \Sigma_U) \propto {\textstyle \prod_{d=1}^{D}} \mathcal{N}(Y_{d}; J(U)\beta, \Sigma_{Y})\cdot  \mathcal{N}(U_{d}; U, \Sigma_U)\cdot \mathcal{N}(U; U_{0}, \Sigma_{U_0})\cdot  \mathcal{N}(\beta; \beta_{0}, \Sigma_{\beta}), \notag
\end{equation}
and the cost in \eqref{eq:kkteiv} is obtained by taking its negative logarithm, yielding
\begin{align*}
    P(\beta, U | \Sigma_U) = &{\textstyle \sum_{d = 1}^{D}} \Big((U - U_{d})^{\top} \Sigma_{U}^{-1} (U - U_{d}) +  (J(U)\beta - Y_{d})^{\top} \Sigma_{Y}^{-1} (J(U)\beta - Y_{d})\Big)  \\  & + (U -U_{0})^{\top} \Sigma_{U_0}^{-1} (U - U_{0}) + (\beta - \beta_{0})^{\top} \Sigma_{\beta}^{-1} (\beta - \beta_{0}).
\end{align*}
We now focus on the case in which the dynamics are linear, the objective features $\phi(\cdot)$ are quadratic, and constraints entering \eqref{eq:optimizationproblem} are polytopic, and derive the Gibbs sampler to build the posterior $p(\beta,\Sigma_{U},U \vert \Sigma_{Y},Y_{[1:D]},U_{[1:D]})$.
\paragraph*{Full conditionals entering the Gibbs sampler} In the set-up mentioned above, $J(U)$ can be re-written as an affine transformation in $U$, i.e., $J(U)\beta = (MU + E)\beta$. The full conditionals entering the posterior are the following:
\begin{enumerate}[label=(\Roman*)] 
    \item $p(\beta|Y_{[1:D]},U)$ as in \eqref{eq:fullbeta}
    \item $p(U|Y_{[1:D]},U_{[1:D]}, \Sigma_U, \beta)$ as in \eqref{eq:fullU}.
    \item $p(\Sigma_U|U_{[1:D]}, U)$ as in \eqref{eq:fullSigmaU}.
\end{enumerate}
We now deploy the properties of conjugate priors reviewed in Section \ref{sec:MCMCprimer} to derive their expressions: the first two follow the first row of Table \ref{tab:conjpriors}, while the third is built according to its second row.
\begin{framed}
(I) Let us specify the Gaussian linear model of interest. Considering the vector $Y_{[1:D]} = [Y_1^{\top},..., Y_D^{\top}]^{\top}$, then we have
\begin{equation}
Y_{[1:D]} = \underbrace{[\mathbb{1}_{D}\otimes (MU+E)]}_{J_{Y_{[1:D]}}}\beta + e_Y, \qquad e_Y \sim \mathcal{N}(0, \underbrace{\mathbb{I}_D\otimes \Sigma_Y}_{\Sigma_{Y_{[1:D]}}}). \notag
\end{equation}
Using the prior $\beta \sim \mathcal{N}(\beta_0,\Sigma_{\beta})$, we obtain
\begin{equation}
\begin{aligned}
& \qquad \qquad \qquad \qquad   p(\beta|Y_{[1:D]},U) = \mathcal{N}(\beta; \mu_{\pi(\beta)}, \Sigma_{\pi(\beta)}), \text{ where }\\
&\begin{cases}
\mu_{\pi(\beta)} =\beta_0 + (\Sigma_{\beta}^{-1} + J_{Y_{[1:D]}}^{\top}\Sigma_{Y_{[1:D]}}^{-1}J_{Y_{[1:D]}})^{-1}J_{Y_{[1:D]}}^{\top}\Sigma_{Y_{[1:D]}}^{-1}(Y_{[1:D]} - J_{Y_{[1:D]}}\beta_0), \  \\
\Sigma_{\pi(\beta)} = (\Sigma_{\beta}^{-1} + J_{Y_{[1:D]}}^{\top}\Sigma_{Y_{[1:D]}}^{-1}J_{Y_{[1:D]}})^{-1}
\end{cases}
\label{eq:fullBetaexp}
\end{aligned}
\end{equation}
\end{framed}

\begin{framed}
 (II) Letting $Y_{[1:D]} = [Y_1^{\top},..., Y_D^{\top}]^{\top}$ and $U_{[1:D]} = [U_1^{\top},..., U_D^{\top}]^{\top}$, and defining $M_{\beta} = \nabla_U(MU\beta)$, the linear Gaussian model reads as
\begin{equation}
\underbrace{\begin{bmatrix}
Y_{[1:D]} \\ U_{[1:D]}
\end{bmatrix}}_{D} = 
\underbrace{\begin{bmatrix}
\mathbb{1}_{D,1}\otimes (E\beta) \\ \mathbb{0}_{mND,1}
\end{bmatrix}}_{\mu_D} + \underbrace{\begin{bmatrix}
\mathbb{1}_D\otimes M_{\beta} \\ \mathbb{I}_{mND,1}
\end{bmatrix}}_{S_D}U + e_D,  \quad e_D \sim \mathcal{N}{\scriptsize\Bigg(\mathbb{0}_{2mND,1}, \ \underbrace{\begin{bmatrix} \mathbb{I}_D \otimes \Sigma_Y & \\ & \Sigma_U  \end{bmatrix}}_{\Sigma_D}\Bigg)} . \notag
\end{equation}
Using the prior $U \sim \mathcal{N}(U_0, \Sigma_{U_0})$, we obtain
\begin{equation}
\begin{aligned}
    & \qquad  p(U|Y_{[1:D]}, U_{[1:D]},\beta, \Sigma_U) = \mathcal{N}(U; \mu_{\pi(U)}, \Sigma_{\pi(U)}), \text{ where }\\ &\begin{cases}
        \mu_{\pi(U)} = U_0 + (\Sigma_{U_0}^{-1} + S_D^{\top}\Sigma_D^{-1}S_D)^{-1}S_D^{\top}\Sigma_D^{-1}(D - S_DU_0 - \mu_D)\\
    \Sigma_{\pi(U)} = (\Sigma_{U_0}^{-1} + S_D^{\top}\Sigma_D^{-1}S_D)^{-1}.
    \end{cases}
    \end{aligned}
    \label{eq:fullUexp}
\end{equation}
\end{framed}

\begin{framed}
(III) Finally, again let $U_{[1:D]} = [U_1^{\top},..., U_D^{\top}]^{\top}$. Then, with the prior $\Sigma_U \sim IW(W_U, m_U)$, we obtain the full conditional for $\Sigma_U$ as
\begin{equation}
p(\Sigma_U|U_{[1:D]}, U)) = IW\Bigg(W_U + \sum_{d=1}^D (U_d - U)(U_d - U)^{\top},  D + m_U\Bigg).
\label{eq:fullSigmaUexp}
\end{equation}
\end{framed}

\paragraph*{Summary of the adopted Gibbs sampler}
The iterative sampling scheme of the Gibbs sampler is summarized in Algorithm \ref{alg:Gibbs}.

\begin{algorithm}[h]
\SetAlgoLined
\KwData{Measurements $Y_{[1:D]}$, $U_{[1:D]}$,  covariance $\Sigma_Y$, number of iterations $N_{MCMC}$}
    \KwResult{samples from $p(U,\beta,\Sigma_U \ | \ Y_{[1:D]}, U_{[1:D]}, \Sigma)Y)$}

\For{$j=1,...,N_{MCMC}$}{
\eIf{$j=1$}{
Set $U(1) = \frac{1}{D}\sum_{d=1}^D U_d$\; 

Set $\beta(1) = KKT(U(1))$ (as defined above \eqref{eq:kktopt})  \;

Set $\Sigma_U = \mathbb{I}_{mN}$ \;
}
{Draw $\beta(j)$ from $p(\beta | U(j-1),  Y_{[1:D]})$ according to \eqref{eq:fullBetaexp}  \; 

Draw $U(j)$ from $p(U | \beta(j),  \Sigma_U(j-1), Y_{[1:D]}, U_{[1:D]})$ according to \eqref{eq:fullUexp} \;

Draw $\Sigma_U(j)$ from $p(\Sigma_U | U(j), U_{[1:D]})$ according to \eqref{eq:fullSigmaUexp} \;
}
}

\caption{Gibbs sampling adopted in Section \ref{subsec:mlwithcovupdate}}
\label{alg:Gibbs}    
\end{algorithm}

Then, to address our inverse optimal control problem stated in \eqref{eq:kkteiv}, we use the last $N_S = 300$ samples in the output of Algorithm \ref{alg:Gibbs} to compute the sample mean for $\Sigma_U$ to be plugged in the cost in \eqref{eq:kkteiv}. Furthermore, the same rationale of Monte Carlo integration is used to compute an estimate for $U$,  which is then utilized to retrieve an initial guess for $\beta$ and provide the initialization for the MPEC solver.

\paragraph*{What if linearity in $J(U)$ is lost}
If the dynamics are nonlinear, then the likelihood yielding \eqref{eq:fullUexp} loses the desired linear dependence on $U$, therefore the full conditional is not Gaussian anymore.  In this case, as hinted at in Section \ref{subsec:mlwithcovupdate}, the sampling strategy is modified by substituting \eqref{eq:fullUexp} with a Metropolis-Hastings step.  In our experiments reported in Sections \ref{sec:bycicle} and \ref{sec:glucose}, we do so by considering a proposal density $q(U(j)\vert U(j-1),\Sigma_U(j-1)) = \mathcal{N}(U(j);U(j-1),a\Sigma_{U}(j-1))$. The scaling value $a$ is chosen to obtain an acceptance rate of approximately 40\%, and is thus set to~$10^{-5}$.

\bibliography{bibliography}

\end{document}